\documentclass{elsart}
\usepackage{amssymb}
\usepackage{graphicx}
\usepackage{longtable}
\usepackage{nicefrac,units}
\usepackage{amsmath}
\usepackage[ngerman,english]{babel}
\usepackage[square,comma,numbers,sort&compress]{natbib}
\usepackage{ifthen}
\newcommand*{\wn}{cm$^{-1}$}

\def\apb{Appl.\  Phys.\  B}

\def\cjc{Can.\ J.\ Chem.\ }

\def\pra{Phys.\ Rev.\ A }
\def\prl{Phys.\ Rev.\ Lett.\ }

\def\oex{Opt.\ Expr.\ }

\def\np{Nat.\ Phot.\ }

\def\jms{J. Mol.\ Spectrosc.\ }
\def\jpb{J. Phys.\ B }

\def\molp{Mol.\ Phys.\ }

\def\rsi{Rev. Scient.\ Instr. }
\def\rmp{Rev. Mod.\ Phys. }

\def\josab{J. Opt.\ Soc.\ Am.\ B }

\def\aaa{Astron.\ Astrophys.\  }

\begin{document}
\selectlanguage{english}
\begin{frontmatter}

\title{Novel techniques in VUV high-resolution spectroscopy}

\author{W. Ubachs}
\ead{w.m.g.ubachs@vu.nl}
\author{$^1$, E. J. Salumbides$^1$,}
\author{K. S. E. Eikema$^1$,}
\author{N. de Oliveira$^2$,}
\author{L. Nahon$^2$}
\address{$^1$ Department of Physics and Astronomy, LaserLaB,
VU University, De Boelelaan 1081, 1081 HV Amsterdam, The Netherlands}
\address{$^2$Synchrotron Soleil, Ormes des Merisiers St Aubin BP 45,
91192 Gif sur Yvette Cedex, France}
\begin{abstract}

Novel VUV sources and techniques for VUV spectroscopy are reviewed. Laser-based VUV sources have been developed
via non-linear upconversion of laser pulses in the nanosecond (ns), the picosecond (ps), and femtosecond (fs) domain, and are applied in high-resolution gas phase spectroscopic studies.
While the ns and ps pulsed laser sources, at Fourier-transform limited bandwidths, are used in wavelength scanning
spectroscopy, the fs laser source is used in a two-pulse time delayed mode.
In addition a Fourier-transform spectrometer for high resolution gas-phase
spectroscopic studies in the VUV is described, exhibiting the multiplex advantage to measure
many resonances simultaneously.

\end{abstract}

%\begin{keyword}

%\PACS 33.20.Ni 33.15.Mt
%\end{keyword}

\end{frontmatter}

\section{Introduction}

In the wavelength region of the vacuum ultraviolet, i.e. the range $50 - 200$ nm, most molecules
exhibit an unstructured continuum spectrum, since the high energy photons probe the level structure
above one or more dissociation limits. However, some important molecules exhibit a spectrum of narrow
lines even at wavelengths as short as 100 nm. The carbon monoxide (CO) molecule has its first dissociation
limit beyond 10 eV, and its spectrum in the range 80-120 nm is characterized by strongly perturbed rovibronic
resonances, where virtually all lines undergo predissociation. These phenomena are of great importance for the
investigation of the chemical dynamics of the interstellar medium, in particular of star-forming regions,
where the photodissociation of CO is the governing dynamical process and CO is, at the same time, the coolant
molecule~\cite{visser2009}. The nitrogen molecule (N$_2$), iso-electronic to CO, has an onset of a dipole-allowed
absorption spectrum at 100 nm, while its spectrum is similarly perturbed and prone to predissociation.
The photoabsorption of N$_2$ is of great importance for the dynamics of the upper layers of the
Earth's atmosphere, and recently
the importance of N$_2$ photo-predissociation for the interstellar medium is increasingly recognized~\cite{li2013}.
The dipole-allowed absorption spectrum of molecular hydrogen (H$_2$) has its strongest lines in the
Lyman and Werner bands in the wavelength range $90 - 115$ nm.
High-resolution spectroscopic studies and high-accuracy wavelength calibration studies
of these sharp H$_2$ resonances has become of specific relevance for studies searching for
a possible variation of the proton-electron mass ratio on a cosmological time
scale~\cite{ubachs2004,reinhold2006}.

In the domain of high-resolution studies of atoms in particular the ground state
of the Helium atom has been the target of many detailed studies over decades.
This two-electron system is amenable for accurate
\emph{ab initio} calculations including quantum electrodynamic (QED) effects.
The accurate determination of the ground state
Lamb shift has been a driving force for the development of VUV-lasers and
VUV-laser spectroscopic techniques since the first
laser excitation of the first resonance line in the Helium atom~\cite{eikema1993}.

\section{Nano-second pulsed VUV sources}

\begin{figure}
\includegraphics[width=0.8\columnwidth]{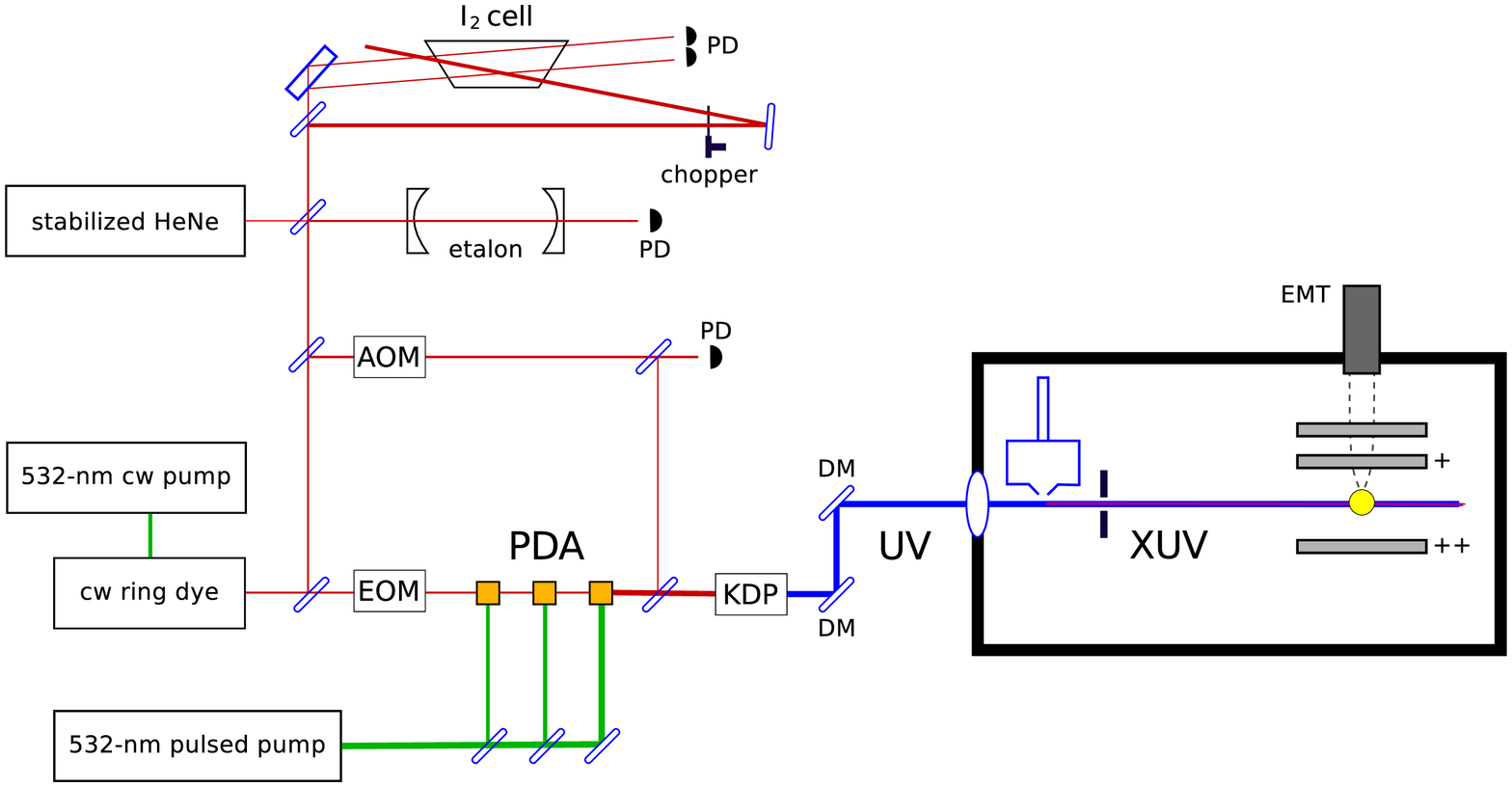}
\caption[]{\small{Experimental setup of the pulsed dye-amplifier based narrowband
and tunable VUV source, in a crossed beam configuration for performing on-line
Doppler reduced spectroscopy. In the upper part the calibration facilities, a HeNe stabilized
etalon and an I$_2$ saturation spectroscopic setup are shown. The box in the lower right corner
shows the differentially pumped vacuum setup for the production of VUV in the first zone,
and the excitation of a perpendicularly crossing molecular beam and the ion extraction,
time-of-flight and detection system. The acousto-optic modulator (AOM) serves to shift the carrier
frequency in the chirp detection scheme. The electro-optic modulator serves to modulate the phase
of the seed-light in order to compensate for chirp effects in the amplifier~\cite{eikema1997}.}}
\label{PDA}
\end{figure}

Techniques of harmonic generation in gases for the production of tunable radiation in the VUV domain have become
well-established over the years. Although resonant sum- and difference-frequency techniques have many
advantages to produce the best intensities and to cover a wide wavelength range~\cite{hollenstein2000}, the
approach of non-resonant third-harmonic generation has certain advantages for studies where absolute frequency calibration is the central concern. The setup displayed in Fig.~\ref{PDA} shows an example
of a narrowband tunable laser system, based on a traveling-wave pulsed three-stage pulsed dye amplifier (PDA),
which is seeded by the output of a continuous wave (CW) ring dye-laser~\cite{ubachs1997}.
By this means laser pulses of Fourier-transform limited
bandwidth are produced, corresponding to the pulse duration (5 ns) of the frequency-doubled Nd:YAG laser employed
for pumping the PDA. The intense PDA pulses (typically 200 mJ) are easily converted to the UV via second harmonic generation in a KDP crystal.
Upon focusing the UV pulses underneath the orifice of a pulsed valve in a jet of xenon, third harmonics are generated, therewith producing VUV pulses at the exact 6$^{th}$ harmonic of the fundamental. When running the ring dye laser in its operating range of $\lambda_f = 560 - 690$ nm, tunable radiation
in the VUV range of $\lambda_{VUV} = 93 - 115$ nm can be produced. In view of the pulses being somewhat shortened
in the non-linear upconversion process, the typical duration is $\sim 3$ ns and the bandwidth in the VUV is about 250 MHz. This bandwidth slightly exceeds the Fourier-bandwidth due to chirp effects in the dye amplifier and harmonic upconversion.

{\bf Figure~\ref{PDA}}

\begin{figure}
\includegraphics[width=0.6\columnwidth]{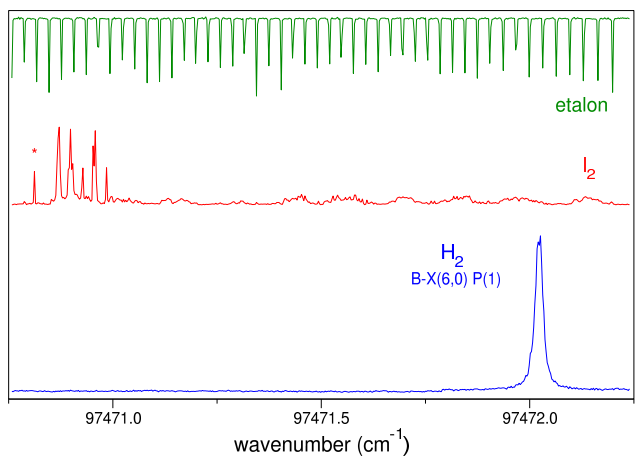}
\caption[]{\small{Measurement of the B-X (6,0) P(1) line of H$_2$,
at 94\,472.029 \wn, recorded by 1 VUV + 1 UV resonance enhanced photoionization.
The absolute calibration is performed against the I$_2$ line marked with an asterisk;
this is the a$_1$ hyperfine component of the P(34) line in the B-X (10,3) band
at 16\,245.13532 \wn~\cite{xu2000}.}}
\label{H2line}
\end{figure}

The main advantage of the system displayed in Fig.~\ref{PDA} is that the VUV output is governed by the exact 6$^{th}$
harmonic of the seed-frequency, a property which may be exploited in the accurate calibration of the
measured resonances in the VUV. The frequency of the (CW) seed-light is accurately calibrated by performing saturated
absorption spectroscopy on I$_2$, for which the individual hyperfine components are known to an accuracy of 1 MHz~\cite{velchev1998,xu2000}. The separation between an I$_2$ resonance at the the fundamental and the VUV-resonances at the 6$^{th}$ harmonic is then bridged in terms of fringes from a stabilized etalon, locked to a
stabilized HeNe-laser. This procedure is illustrated in Fig.~\ref{H2line}, showing a recording of a specific line
in the Lyman band system of the H$_2$ molecule. As in many experiments employing laser-based VUV radiation, a measurement scheme is chosen employing 1 VUV + 1 UV resonance-enhanced two-photon ionization. In this scheme the VUV radiation resonantly excites a high-lying state in the atomic/molecular system, which is then further ionized
by the spatially and temporally overlapping UV beam, i.e. the leftover beam from the harmonic conversion in the
gas jet.
The error budget of VUV absolute frequency measurements contains contributions of statistics, and of systematic effects related to the AC Stark effect (induced primarily by the UV beam), and the residual Doppler effect due to non-perfect perpendicular alignment of the laser beams with respect to the collimated molecular beam. An additional source of uncertainty is related to possible chirp effects in the pulsed laser beams, as a result from time-dependent gain in the dye amplifiers. Experimental methods have been developed to determine the effect of this chirp on the
measured transition frequency, and even to compensate the chirp effect by adapting the phase of the incident seed-light by an electro-optic modulator~\cite{eikema1997}.

{\bf Figure~\ref{H2line}}

A comprehensive study has been performed on the VUV absorption spectrum of H$_2$ with the focus on accurate wavelength calibration for all the lines in the Lyman (B$^1\Sigma_u^+$ - X$^1\Sigma_g^+$) and Werner
(C$^1\Pi_u$ -  X$^1\Sigma_g^+$) bands~\cite{philip2004}. Similar calibrations were performed on the HD isotopomer~\cite{hollenstein2006,ivanov2008}. These studies provide a database of accurate absolute wavelengths
for all H$_2$ and HD absorption lines, at an accuracy of $\Delta\lambda/\lambda = 4 \times 10^{-8}$. This database
forms an ingredient for a study of possibly varying constants on a cosmological time scale. Thereby use is made of the equation:
\begin{equation}
 \frac {\lambda_i^z}{\lambda_i^0} =  (1+z_\mathrm{abs})(1+ \frac{\Delta \mu}{ \mu}K_i)
 \label{drift-eq}
\end{equation}
making a comparison of the wavelengths $\lambda_i^z$ as obtained from astronomical observations at high redshift $z$ to the wavelengths $\lambda_i^0$ in the database (at zero redshift $z=0$), where $z_{abs}$
is the overall redshift of an absorbing
galaxy with observable H$_2$ abundance in the line of sight toward a quasar source.
The final factor on the right
hand side of the equation represents the possible effect of a drifting proton-electron mass ratio $\mu = m_p/m_e$,
with $\Delta\mu/\mu$ the relative change in its value.
For a comparison based on Eq.~(\ref{drift-eq}) a calculation of sensitivity coefficients $K_i$ is required.
Their values can be calculated from the known level structure of the H$_2$ molecule,
by semi-empirical methods via~\cite{ubachs2007}:
\begin{equation}
K_i=\frac{d\ln\lambda_i}{d\ln\mu}
\end{equation}
Data on high-redshift H$_2$ absorptions are derived from astronomical observations using the largest optical dishes and high-resolution spectrometers (the Very Large Telescope with the UVES spectrometer, and the Keck telescope with the HIRES spectrometer). A comparison to accurate laboratory wavelengths, and invoking the calculated $K_i$ sensitivity constants then leads to estimates on a possible variation
of the proton-electron mass ratio. The most detailed studies~\cite{king2008,malec2010,weerdenburg2011} yield the
overall result that the proton-electron mass ratio has not changed by more than 0.001 \%, or
$|\Delta\mu/\mu| < 10^{-5}$ for redshifts of $z=2-3.5$. This corresponds to look-back times of 10-12 billion years.
This important result relies on the accurate laboratory calibration of H$_2$ and HD spectroscopic lines in the VUV.

\begin{figure}
\includegraphics[width=0.5\columnwidth]{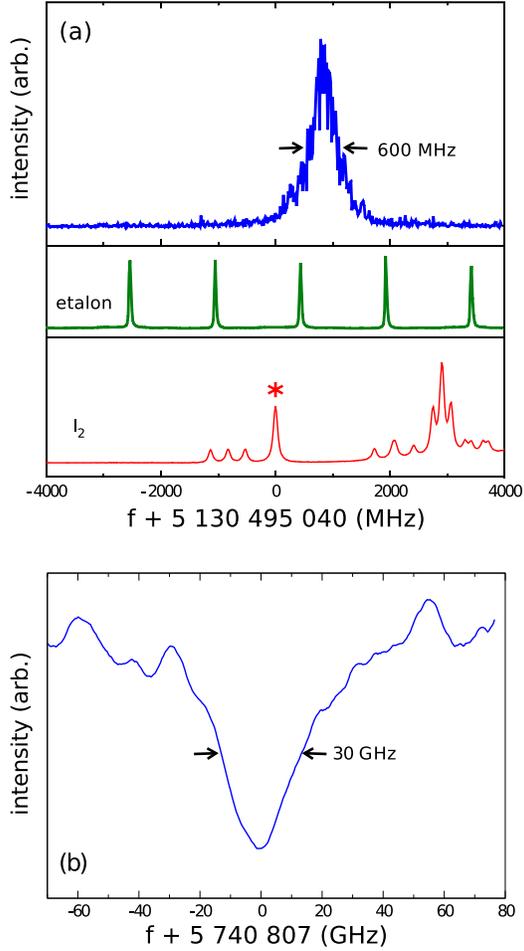}
\caption[]{\small{Excitation resonances of he atomic lines. (a) 1 VUV + 1 UV photoionization of the
$1s^2\,^1 \rm{S}_0$ - $1s2p\,^1 \rm{P}_1$ line of helium at $54.8$ nm produced with the ns-PDA laser system upon
5$^{th}$ harmonic conversion at a linewidth of 600 MHz;
(b) Absorption spectrum of the $1s^2\,^1 \rm{S}_0$ - $1s4p\,^1 \rm{P}_1$ line of helium at $52.2$ nm, recorded with the picosecond VUV laser system as shown in Fig.~\ref{Pico}, at a linewidth of 30 GHz.
}}
\label{He-resonances}
\end{figure}

{\bf Figure~\ref{He-resonances}}

Harmonic generation in the perturbative regime, hence at power densities of $< 5 \times 10^{12}$  W/cm$^2$,
usually is restricted to $3^{rd}$ harmonics. However, with the high-power frequency-doubled PDA-laser system
of Fig.~\ref{PDA} small amounts of $5^{th}$ harmonics could be produced, therewith producing the narrowest
bandwidth pulses at wavelengths as short as 58 nm~\cite{eikema1996}. Based on such a VUV system the $1s^2\,^1 \rm{S}_0$ - $1s2p\,^1 \rm{P}_1$ resonance line of the helium atom could be excited at a resonance width of 600 MHz (see Fig.~\ref{He-resonances}).
Careful analysis of systematic effects, in particular the chirp-effects in the PDA-system and in the harmonic
conversion process, allow for an experimental accuracy on the resonance line of $\Delta\lambda/\lambda < 10^{-8}$ and a determination of the Lamb shift in the He ground state at an accuracy of 45 MHz~\cite{eikema1997}.

\section{Pico-second pulsed VUV sources}

\begin{figure*}
\includegraphics[width=\columnwidth]{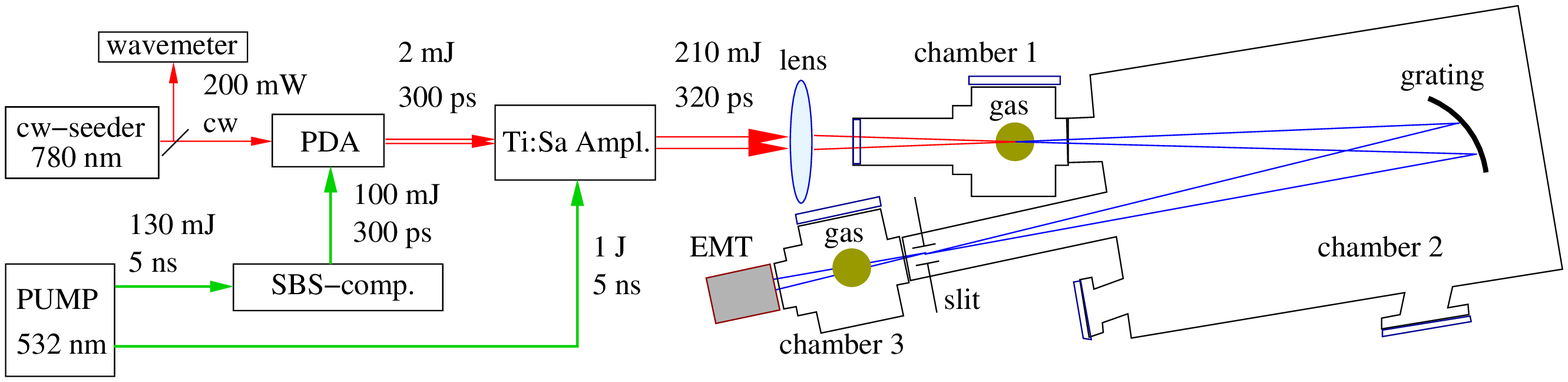}
\caption[]{\small{\emph{Two-column figure.} Setup of the VUV source for pulses in the 300 ps time domain. A pulsed dye amplifier (PDA) is injection seeded by the output of a CW titanium-sapphire (Ti:Sa) laser at $\lambda = 760 - 830$ nm and pumped by the SBS compressed output of a frequency-doubled Nd:YAG laser. These pulses are further amplified in a bowtie amplifier with Ti:Sa crystals. The infrared laser pulses produce, upon focusing at $f=20$ cm, power densities of $> 10^{13}$ W/cm$^2$, sufficient to reach the plateau for generating high harmonics into the wavelength window of
$\lambda = 50$ nm. Figure reproduced from Ref.~\cite{brandi2003b}.}}
\label{Pico}
\end{figure*}

At incident power densities of $> 10^{13}$  W/cm$^2$ a sequence of harmonics can be produced,
known as the plateau region~\cite{huillier1983}. In order to reach the plateau threshold with laser pulses as
narrow as possible in the frequency domain, a laser system was built generating high-power pulses at 300 ps
duration~\cite{brandi2003a}.
A combination of techniques was employed as shown in the setup of Fig.~\ref{Pico}. Powerful pump pulses at 300 ps duration were generated by stimulated Brillouin scattering in a compact configuration using liquid methanol as the
nonlinear medium~\cite{neshev1999}. These pulses pumped a pulsed-dye amplifier, injection seeded by the output
of a CW Ti:Sa laser, running on infrared dyes with typical excited state lifetimes of 300 ps.
This allows for producing
pulses of 300 ps, tunable in the range 700-850 nm at a repetition rate of 10 Hz. Further amplification in a
Ti:Sa preamplifier and a seven-pass Ti:Sa bow-tie amplifier resulted in pulses of $> 250$ mJ/pulse at a
Fourier-transform limited bandwidth of 1.5 GHz with a Fourier-product of $\Delta\nu \times \Delta\tau =0.45$.
After focusing the pulses with a lens with a focal distance of $f=20$ cm, harmonics up to $15^{th}$ were generated~\cite{brandi2003b}.
In this process chirp phenomena cause a rather large broadening of the bandwidth of the pulses in the VUV~\cite{brandi2006}.
This is the result of blue-shift effects from ionization in the early part of the pulse, and red-shift effects due to the plasma expansion in the trailing part of the pulses. Moreover, the combination of pulse duration and peak intensity causes the interaction
volume to become fully ionized already during the front part of the pulse.

{\bf Figure~\ref{Pico}}

This VUV picosecond source was employed in a spectral recording
of the He $1s^2\,^1 \rm{S}_0$ - $1s4p\,^1 \rm{P}_1$ resonance line at 52.2 nm, as shown in Fig.~\ref{He-resonances}.
The linewidth in the spectrum, amounting to 30 GHz, is mainly the result of the chirp effects in the harmonic generation process. This is a limiting factor to the achievable resolution for this design of a VUV source.

\section{Femtosecond VUV pulses for high resolution spectroscopy}

The ultra-short pulses from a femtosecond laser are seemingly not useful for spectroscopic studies in view of the Fourier-principle, dictating that such pulses exhibit an almost white spectrum. However, if the phase relationships
between consecutive pulses are controlled and used in combination, the perspective alters drastically. This is
ultimately reached in a frequency comb laser where full control is achieved for an infinite pulse train, thus
delivering a mode spectrum:
\begin{equation}
 f_n = n f_{rep} + f_{CEO}
 \label{comb-modes}
\end{equation}
where $f_{rep}$ is the repetition frequency of the modelocked frequency comb, $f_{CEO}$ the carrier-envelope offset frequency, and $n$ a large integer number. Both $f_{rep}$ and $f_{CEO}$ can be measured or locked against an atomic clock reference so that all modes of the comb laser are known with high precision~\cite{hansch-nobel,hall-nobel}.

The production of high-order harmonics in the plateau region can be easily achieved
for the ultrashort pulses from a frequency comb laser, in particular when a single pulse-pair is selected from the pulse train and subsequently amplified in a non-collinear optical parametric amplifier~\cite{kandula2008}. Under the condition that no extra phases are imposed upon the laser pulses in the amplification process nor in the harmonic conversion process, the resulting comb spectrum after conversion into the VUV is:
\begin{equation}
 f_m = m f_{rep} + q f_{CEO}
 \label{vuv-modes}
\end{equation}
where $m$ is again an integer and $q$ is the integer harmonic order. The phase build-up in the amplifiers can be measured by interferometric techniques~\cite{kandula2008}, while the phases acquired in harmonic conversion can be
measured as well~\cite{kandula2011}, and are therefore left out of consideration. It is noted however, that only
the \emph{difference} in phase build-up between consecutive pulses under consideration, $\Delta\psi$, is of relevance.

\begin{figure}
\includegraphics[width=0.8\columnwidth]{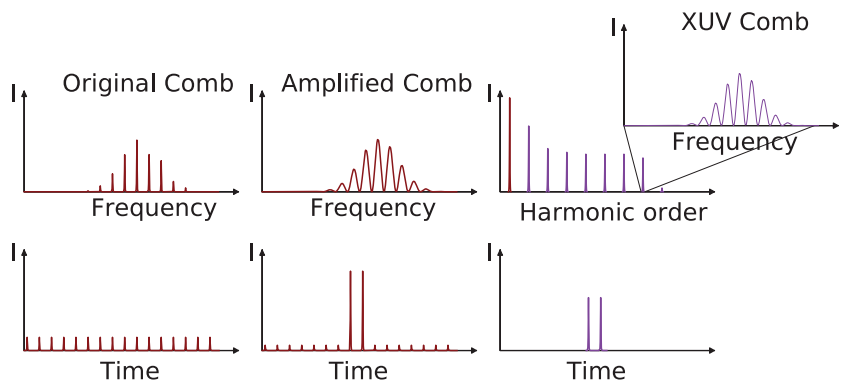}
\caption[]{\small{Sequence of manipulation of the comb structure from a frequency comb,
with the full pulse train emanating from the comb laser, the amplification of a pair
of two pulses in a non-collinear optical parametric amplifier, and the conversion into harmonics
in the plateau region. The 15$^{th}$ harmonic, to be used in the excitation of the helium atom,
exhibits a cosine-modulated mode spectrum with period $f_{rep}$.}}
\label{XUV-steps}
\end{figure}

In case two consecutive pulses are selected from a full pulse train emanating from a frequency comb laser the mode spectrum stays the same, except for the fact that the modes are no longer sharp spikes, but are diffused into a cosine-modulated mode spectrum. Upon harmonic conversion of the double-pulse structure the same mode structure as in Eq.~(\ref{vuv-modes}) is retained~\cite{kandula2008} but again as a cosine modulation. This sequence of manipulative steps on the comb structure and the spectrum is highlighted in Fig.~\ref{XUV-steps}.

{\bf Figure~\ref{XUV-steps}}

\begin{figure}
\includegraphics[width=0.8\columnwidth]{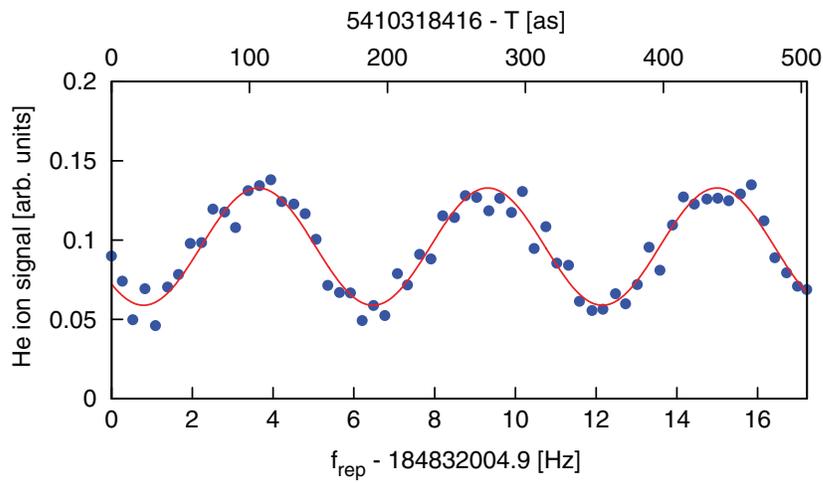}
\caption[]{\small{Two-pulse excitation probability on the $1s^2\,^1 \rm{S}_0 - 1s5p\,^1 \rm{P}_1$
transition in Helium. The lower $x$-axis represents the scan of the repetition rate of the frequency comb laser,
while on the upper axis the delay between the two pulses is shown in attoseconds. Figure from Ref.~\cite{kandula2011}.}}
\label{FC-Comb}
\end{figure}

When using such double pulses for measuring a transition at frequency $f_{tr}$ in an atomic system, one has to scan the time separation $T$ between the two pulses so that an excitation spectrum is obtained of the form:
\begin{equation}
 S(T) \propto \cos[2\pi f_{tr}T]
 \label{excitation}
\end{equation}
where it is assumed that additional phase difference effects are controlled and may be neglected, so that $\Delta\psi(f_{tr})=0$~\cite{kandula2011}. The resulting cosine-modulated excitation spectrum is displayed in
Fig.~\ref{FC-Comb} representing a measurement of the $1s^2\,^1 \rm{S}_0$ - $1s5p\,^1 \rm{P}_1$
transition in helium, using the 15$^{th}$ harmonic of a Ti:Sa frequency comb laser at a wavelength of 51 nm in the VUV.

{\bf Figure~\ref{FC-Comb}}

This form of two-pulse delayed time-domain spectroscopy may be compared to Ramsey's separated zone
oscillatory field spectroscopy, where now the separation is in the time domain rather than in the spatial domain.
In particular when using all pulses from a frequency comb laser it is usually referred to as \emph{direct frequency comb spectroscopy}~\cite{marian2004}, and as in all
absolute frequency measurements employing frequency comb lasers the central mode number, in this case the integer $m$
of Eq.~(\ref{vuv-modes}), must be determined. This is accomplished by performing the measurement for a number
of repetition rates $f_{rep}$ of the laser; for this purpose the cavity of the comb laser was redesigned to
allow for settings between 100 MHz and 185 MHz~\cite{kandula2010}. This study results in an accurate value
of the ionization potential of the helium atom as well as a determination of the He ground state (at the $10^{-9}$
accuracy level), including a value for the Lamb shift of the He ground state as accurate as 6 MHz~\cite{kandula2010,kandula2011}.

\section{VUV Fourier-transform spectroscopy}

From the perspective of non-laser based spectroscopies the technique of Fourier-transform (FT) spectroscopy has
been been known as an important tool combining high resolving power and accurate intrinsic wavelength calibration.
Moreover, the interferometric technique exhibits a multiplex advantage. The entire spectrum of the light source
employed is covered at once in the spectral decomposition obtained by Fourier-transforming  an interferogram.
While FT-spectroscopy is widely used in the infrared domain as an analytical sensing tool, it is used in the optical
domain and even in the VUV domain down to wavelengths of 140 nm~\cite{thorne1998}, but operation in all cases depends
on a beam-splitter as a crucial optical element of the interferometer.

\begin{figure}
\includegraphics[width=\columnwidth]{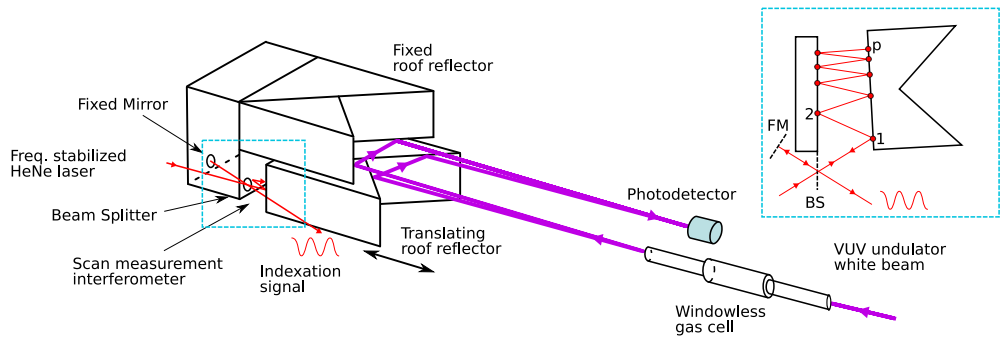}
\caption[]{\small{\emph{Two-column figure.} Experimental layout of the wavefront-division interferometer
used at the DESIRS beam line at Soleil with the roof-shaped VUV reflecting surfaces in the path
to produce the interferences at the detector (PD). In the zoomed-in part (upper right corner) the control of the beampath difference, established by a HeNe
laser beam reflecting multiple times from the rear of the traveling optic, is shown. Figure from Ref.~\cite{ivanov2010}.}}
\label{Rooftop}
\end{figure}

A recent development in VUV spectroscopy is the operation of a wave-front division setup to
generate an interferogram between two spatial parts of a propagating beam at VUV wavelengths, without
the use of transmitting materials.
The scanning interferometer, shown in some detail in Fig.~\ref{Rooftop}, exploits the delay
imposed by two rooftop reflectors shifted with respect to each other. The stability and step-increments of the reflector displacement of the spatial scanning instrument should be kept under extreme control, i.e. within a fraction of the wavelength, in order to generate the interferograms. This is accomplished by probing the motion
of one of the roof-shaped reflectors using a HeNe laser control interferometer.
Such a FT-VUV device was developed at the Soleil synchrotron, where it is operated at the
beam-line DESIRS at wavelengths in the range 40-200 nm~\cite{oliveira2011}.

{\bf Figure~\ref{Rooftop}}

The output of the undulator-based beam line delivers a beam of sufficient transverse spatial coherence to operate
in wave-front division. A bell-shaped output spectrum of typically 7\% bandwidth against the continuum background
with sufficient power allows for the recording of interferograms on a VUV silicon photodiode, even in case of the
lossy reflections on the roof-shaped reflectors.

\begin{figure}
\includegraphics[width=0.7\columnwidth]{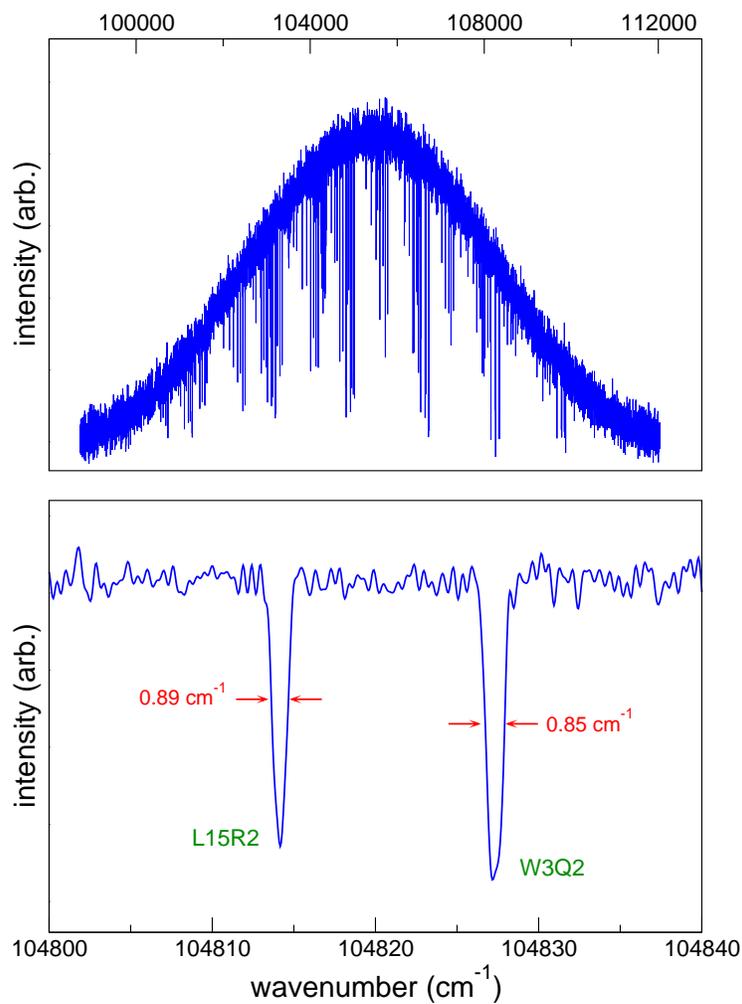}
\caption[]{\small{Fourier-transform VUV absorption spectrum of the HD molecule at room temperature measured at the DESIRS beamline at Soleil. The top panel displays the bell-shaped curve associated with a single setting of the undulator. The bottom spectrum is a zoom-in displaying a line in the Lyman band (L15R2) and a line in the Werner band (W3Q2) at the Doppler width of 0.85 \wn.}}
\label{HD-FT}
\end{figure}

As an application example, a spectrum of high-resolution study of the HD molecule is shown in Fig.~\ref{HD-FT}~\cite{ivanov2010}. An aspect of the FT-VUV spectrometer is that only linear \emph{absorption}
spectra can be recorded, imposing the disadvantage of Doppler broadening as a limitation. In the spectrum
of Fig.~\ref{HD-FT}, taken at room temperature, the Doppler width is the dominant factor in the linewidth
amounting to 0.85 \wn. Recently, also static gas spectra with liquid-N$_2$ and liquid-He cooling of the windowless
flow cell were recorded to reduce the linewidth. In addition the operation of the novel FT-VUV instrument
was also demonstrated in a gas-jet configuration thereby obtaining a linewidth of 0.2 \wn\ in a spectrum of D$_2$~\cite{dickenson2011}.

{\bf Figure~\ref{HD-FT}}

\section*{Acknowledgment}

The authors wish to thank J. Bagdonaite, F. Brandi, P. C. Cacciani, G.D. Dickenson, C. Gohle, A. N. Heays, W. Hogervorst, U. Hollenstein, T. I. Ivanov, D. Joyeux, D. Z. Kandula, C. A. de Lange, M. T. Murphy, M. L. Niu, T. J. Pinkert, J. Philip, E. Reinhold, A. Renault, M. Roudjane, W. Vassen, M. O. Vieitez, S. Witte, A. L. Wolf, and R. T. Zinkstok for their involvement in the projects.
Scientists and staff of the SOLEIL synchrotron are thanked for their support.

\newpage

\clearpage

\clearpage

\clearpage

\clearpage

\clearpage

\clearpage

\clearpage

\end{document}